\documentclass[11pt]{article}

\usepackage[top=0.5in, bottom=0.5in, left=0.5in, right=0.5in]{geometry}
\usepackage{authblk}
\usepackage{hyperref}
\usepackage[utf8]{inputenc}
\usepackage{amsmath}
\usepackage{amsfonts}
\usepackage{amssymb}
\usepackage{siunitx}
\usepackage{graphicx}
\usepackage{subcaption}
\usepackage{float}
\usepackage[nottoc,numbib]{tocbibind}
\usepackage{cite}

\usepackage{setspace}
\usepackage{textcomp}
\usepackage[export]{adjustbox}
\usepackage{longtable}
\usepackage{multirow}
\usepackage{multicol}
\usepackage{xcolor}
\usepackage{pifont}
\usepackage[T1]{fontenc}
\usepackage[justification=centering]{caption}
\usepackage{tikz}
\def\checkmark{\tikz\fill[scale=0.4](0,.35) -- (.25,0) -- (1,.7) -- (.25,.15) -- cycle;} 
\usepackage{makecell}

\title{Fog Computing Resource Management: A Comprehensive Survey on Architectures, Machine Learning Techniques,Tools and Datasets}
\author{ABC, XYZ}

\providecommand{\keywords}[1]
{
  \small	
  \textbf{\textit{Keywords---}} #1
}

\title{A Review of Resource Management in Fog Computing: Machine Learning Perspective}
\author{Muhammad Fahimullah$^{1}$, Shohreh Ahvar$^{1}$, and Maria Trocan$^{1}$  \\
        \small $^{1}$Institut Supérieur d’Électronique de Paris ISEP, Paris, France; \\ muhammad.fahimullah@ext.isep.fr,shohreh.ahvar@isep.fr,  maria.trocan@isep.fr \\
        
}

\makeatletter
\let\inserttitle\@title
\let\insertauthor\@author
\makeatother

\begin{document}

\begin{center}
  \LARGE{\inserttitle}

  \Large{\insertauthor}
\end{center}

\begin{abstract}
    Fog computing becomes a promising technology to process user's requests near the proximity of users to reduce response time for latency-sensitive requests. Despite its advantages, the properties such as resource heterogeneity and limitations, and its dynamic and unpredictable nature greatly reduce the efficiency of fog computing. Therefore, predicting the dynamic behavior of the fog and managing resources accordingly is of utmost importance. In this work, we provide a review of machine learning-based predictive resource management approaches in a fog environment. Resource management is classified into six sub-areas: resource provisioning, application placement, scheduling, resource allocation, task offloading, and load balancing. Reviewed resource management approaches are analyzed based on the objective metrics, tools, datasets, and utilized techniques.
\end{abstract}

\section{Introduction}
The concept of Fog Computing (FC) has emerged recently and is presented in 2012 \cite{bonomi2012fog}. This emergent provides architecture between the cloud and end devices by enabling storage, processing, and data management capabilities near the proximity of users. The configuration, control, data management, and processing of tasks not only takes place on the cloud but all the way from end devices to the cloud \cite{yousefpour2019all,habibi2020fog}. Furthermore, the architecture of FC in the middle of cloud and end devices allows less operational cost and reduces latency, power consumption and network traffic \cite{manasrah2019optimized}. According to the National Institute of Standards and Technology (NIST), \cite{iorga2017nist}, some of the other important characteristics of fog computing are low latency, heterogeneity, geographical distribution, interoperability, federation, and real-time interactions. 

The overall view of FC can be classified into three main components infrastructure, application, and platform \cite{naha2018fog}. Where infrastructure, relates to infrastructure requirements, communication requirements, and fog devices. Similarly, the application relates to application requirements, user requirements, and application modeling. Lastly, the platform relates to Resource Management (RM), security and privacy, multi-tenancy (container or virtualized based), and service requirements. 

\begin{figure}[ht]
\includegraphics[width=0.7\textwidth]{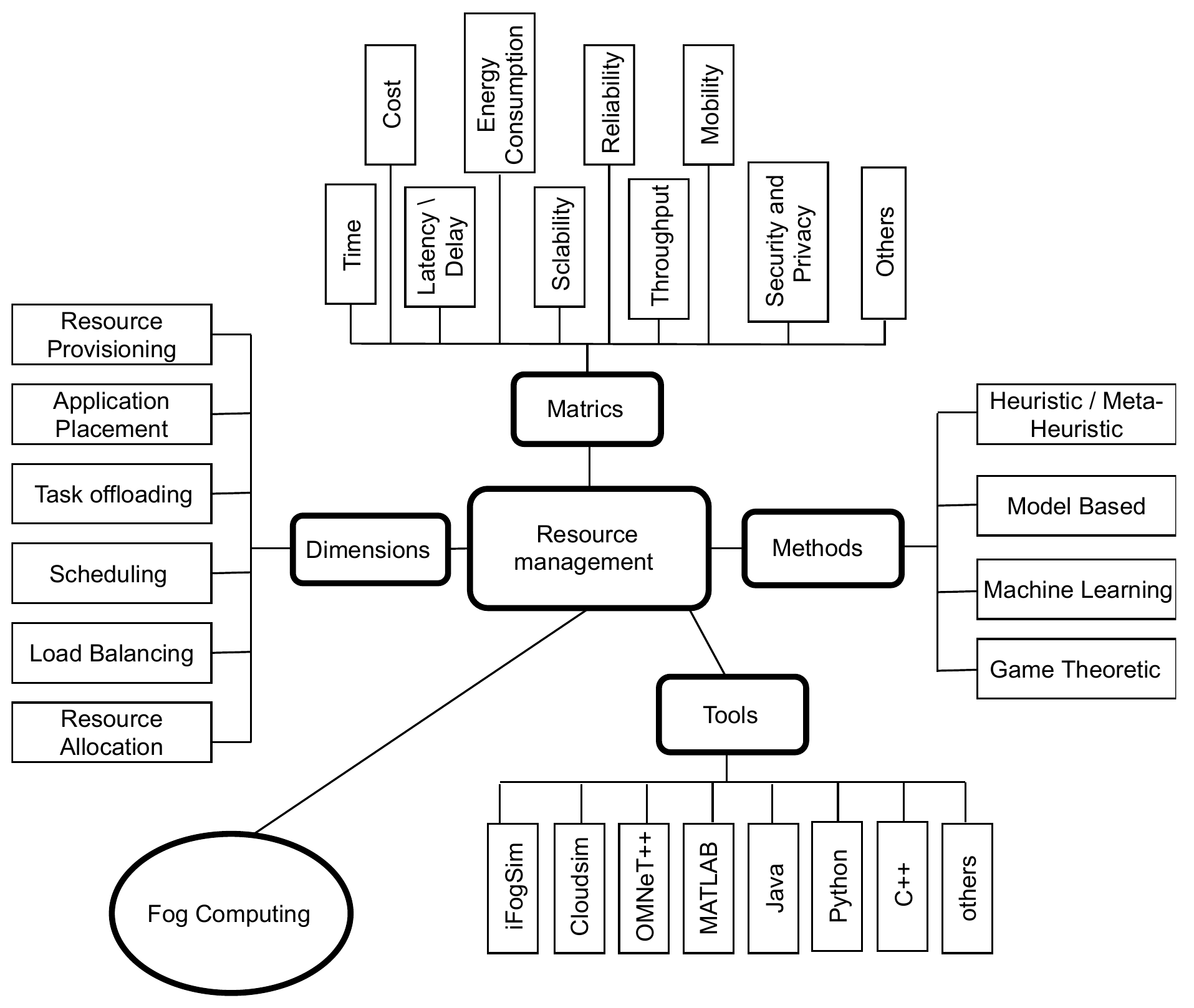}
\caption{Resource Management Aspects} \label{Aspects}
\end{figure}

Although FC provides several benefits, the computational and storage capabilities of devices at the fog layer are comparatively resource constrained compared to the cloud. Therefore, in order to efficiently utilize the fog resources, proper RM is one of the major and challenging issues to be taken into account in FC \cite{ghobaei2020resource}.
Several classifications of RM in fog computing can be found in literature \cite{kansal2022classification,ghobaei2020resource,ahvar2021next,shakarami2022resource}. Based on the literature review, we concluded the generic classification of RM into Six main categories such as resource provisioning, application placement, scheduling, resource allocation, task offloading, and load balancing. These classifications are explained in detail in Section 3. Each of these dimensions is used in literature for achieving different performance metric goals through various methods and tools as shown in Figure \ref{Aspects}.
Some of the important performance matrices in FC are cost, latency\textbackslash delay, energy consumption, scalability, reliability, throughput, mobility, security, and privacy \cite{ghobaei2020resource,kansal2022classification,alsadie2022resource,bukhari2022intelligent}. However, to achieve these performance matrices different methods have been applied in all dimensions of RM. These methods can be classified into  heuristic\textbackslash meta-heuristics, model-based (e.g., Approximation, Markov-based), Machine Learning (ML), and game theoretic \cite{ghobaei2020resource,shakarami2022resource,alsadie2022resource}.

 \begin{table}[ht]
\centering
\caption{Research Questions}
\label{research_questions}
\def\arraystretch{1.2}
\resizebox{\textwidth}{!}{%
\begin{tabular}{l|l}
\hline
\textbf{Category}                       & \textbf{Research Questions}                                                            \\ \hline
Resource Management & How resource management can be classified into sub-areas?                              \\ \hline
Machine Learning Techniques & What are the different ML methods used? \\ \hline
Performance matrices  & What are the different performance matrices considered? \\ \hline
Datasets                                      & Which type of data have been considered for evaluation?       \\ \hline
\multirow{2}{*}{Simulation Tools}            & Which simulation tools have been used?                         \\ \cline{2-2} 
                                    & What sort of simulation tool will be suitable for a certain type of problem?             \\ \hline
\end{tabular}%
}
\end{table}

Nowadays, having FC and Cloud Computing for around a decade and having a record of data, the ML-based methods are current methods in recently proposed solutions. However, the complexity of such methods is the parameter to consider in applying predictive solutions. In this work, we analyze the ML-based methods used for achieving various performance metrics through different tools in RM. 
The research questions we aim to find answers to are mentioned in Table \ref{research_questions}.

The rest of the work is organized as follows. In section 2, we provide a brief overview of some of the related surveys in RM dimensions. Next in section 3, we explain different dimensions of RM along with an overview of ML works, mainly considering the objective performance matrices, algorithms considered, and tools used for evaluation. In the next section, we summarize the works of ML in RM and provided concluding remarks. Lastly, concluding the work with the section conclusion. 

\section{Related Works}
In order to understand the classification of RM in FC, the work in \cite{kansal2022classification}, classified RM into nine sub-categories such as resource scheduling, resource orchestration, resource estimation,  resource allocation, application placement, resource provisioning, resource discovery, load balancing, and task offloading.
On the other hand, the work in \cite{ghobaei2020resource} classified RM into six dimensions: resource provisioning, task offloading, resource scheduling, load balancing, application placement, and resource allocation. Furthermore, this work evaluated the existing literature in the domain of RM on the basis of performance metrics, case studies, utilized techniques, and evaluation tools. However, this work contains a very brief discussion on the architectural perspective of FC. Furthermore, their work has less or negligible literature on the ML aspect of RM. Another study in \cite{ahvar2021next} also provided a review on the use of Software Defined Networks (SDN) in different aspects of RM as per classification of \cite{ghobaei2020resource}. In  addition, the work in \cite{10.1145/3486221}, aimed to contribute to the definition and scope of the orchestration. The study also provided a generic architecture for the orchestration. This work considered the RM as the synonyms for the orchestration and defined the sub-processes of RM such as resource offloading, provisioning, placement, scheduling, allocation, etc. Furthermore, the work provided a very short overview of existing architectures.  
Based on the classification stated in the aforementioned literature, we classified the RM into six main dimensions as shown in Fig \ref{Aspects}.

\begin{table}[ht]
\centering
\caption{Related Surveys}
\label{related_work}
\def\arraystretch{1.5}
\resizebox{\textwidth}{!}{%
\begin{tabular}{cccccccccccc}
\hline
\multicolumn{1}{|c|}{\multirow{2}{*}{\textbf{Ref.}}} &
  \multicolumn{6}{c|}{\textbf{Resource   Management Areas}} &
  \multicolumn{5}{c|}{\textbf{Analysis \& discussion}}\\ \cline{2-12}
\multicolumn{1}{|c|}{} &
  
  \multicolumn{1}{c|}{\textbf{\makecell{RP}}} &
  \multicolumn{1}{c|}{\textbf{\makecell{AP}}} &
  \multicolumn{1}{c|}{\textbf{Scheduling}} &
  \multicolumn{1}{c|}{\textbf{\makecell{RA}}} &
  \multicolumn{1}{c|}{\textbf{\makecell{LB}}} &
  \multicolumn{1}{c|}{\textbf{Offloading}} &
  \multicolumn{1}{c|}{\textbf{Architecture}} &
  \multicolumn{1}{c|}{\textbf{ML}} &
  \multicolumn{1}{c|}{\textbf{Matrices}} &
  \multicolumn{1}{c|}{\textbf{Simulation Tools}} &
  \multicolumn{1}{c|}{\textbf{Dataset}}\\ \hline

\multicolumn{1}{c}{ \cite{ghobaei2020resource}} &
  \multicolumn{1}{c}{\checkmark} &
  \multicolumn{1}{c}{\checkmark} &
  \multicolumn{1}{c}{\checkmark} &
  \multicolumn{1}{c}{\checkmark} &
  \multicolumn{1}{c}{\checkmark} &
  \multicolumn{1}{c}{\checkmark} &
  \multicolumn{1}{c}{$\partial$} &
  \multicolumn{1}{c}{$\partial$} &
  \multicolumn{1}{c}{$\partial$}  &
  \multicolumn{1}{c}{$\partial$} &
  \multicolumn{1}{c}{} \\ \hline
  
\multicolumn{1}{c}{ \cite{kansal2022classification}} &
  \multicolumn{1}{c}{\checkmark} &
  \multicolumn{1}{c}{\checkmark} &
  \multicolumn{1}{c}{\checkmark} &
  \multicolumn{1}{c}{\checkmark} &
  \multicolumn{1}{c}{\checkmark} &
  \multicolumn{1}{c}{\checkmark} &
  \multicolumn{1}{c}{$\partial$} &
  \multicolumn{1}{c}{$\partial$} &
  \multicolumn{1}{c}{$\partial$}  &
  \multicolumn{1}{c}{$\partial$} &
  \multicolumn{1}{c}{} \\ \hline

\multicolumn{1}{c}{ \cite{rahimikhanghah2021resource}} &
  \multicolumn{1}{c}{} &
  \multicolumn{1}{c}{} &
  \multicolumn{1}{c}{\checkmark} &
  \multicolumn{1}{c}{} &
  \multicolumn{1}{c}{} &
  \multicolumn{1}{c}{} &
  \multicolumn{1}{c}{$\partial$} &
  \multicolumn{1}{c}{$\partial$} &
  \multicolumn{1}{c}{$\partial$}  &
  \multicolumn{1}{c}{$\partial$} &
  \multicolumn{1}{c}{$\partial$} \\ \hline

\multicolumn{1}{c}{\cite{alizadeh2020task} } &
  \multicolumn{1}{c}{} &
  \multicolumn{1}{c}{} &
  \multicolumn{1}{c}{\checkmark} &
  \multicolumn{1}{c}{} &
  \multicolumn{1}{c}{} &
  \multicolumn{1}{c}{} &
  \multicolumn{1}{c}{$\partial$} &
  \multicolumn{1}{c}{$\partial$} &
  \multicolumn{1}{c}{\checkmark}  &
  \multicolumn{1}{c}{$\partial$} &
  \multicolumn{1}{c}{} \\\hline
  
  \multicolumn{1}{c}{\cite{kaur2021systematic} } &
  \multicolumn{1}{c}{} &
  \multicolumn{1}{c}{} &
  \multicolumn{1}{c}{\checkmark} &
  \multicolumn{1}{c}{} &
  \multicolumn{1}{c}{} &
  \multicolumn{1}{c}{} &
  \multicolumn{1}{c}{$\partial$} &
  \multicolumn{1}{c}{$\partial$} &
  \multicolumn{1}{c}{\checkmark}  &
  \multicolumn{1}{c}{\checkmark} &
  \multicolumn{1}{c}{} \\\hline
  
  \multicolumn{1}{c}{\cite{singh2022towards} } &
  \multicolumn{1}{c}{} &
  \multicolumn{1}{c}{} &
  \multicolumn{1}{c}{\checkmark} &
  \multicolumn{1}{c}{} &
  \multicolumn{1}{c}{} &
  \multicolumn{1}{c}{} &
  \multicolumn{1}{c}{$\partial$} &
  \multicolumn{1}{c}{} &
  \multicolumn{1}{c}{\checkmark}  &
  \multicolumn{1}{c}{\checkmark} &
  \multicolumn{1}{c}{} \\\hline
    
  \multicolumn{1}{c}{\cite{nayeri2021application} } &
  \multicolumn{1}{c}{} &
  \multicolumn{1}{c}{\checkmark} &
  \multicolumn{1}{c}{} &
  \multicolumn{1}{c}{} &
  \multicolumn{1}{c}{} &
  \multicolumn{1}{c}{} &
  \multicolumn{1}{c}{$\partial$} &
  \multicolumn{1}{c}{\checkmark} &
  \multicolumn{1}{c}{$\partial$}  &
  \multicolumn{1}{c}{$\partial$} &
  \multicolumn{1}{c}{} \\\hline
  
  \multicolumn{1}{c}{\cite{mahmud2020application} } &
  \multicolumn{1}{c}{} &
  \multicolumn{1}{c}{\checkmark} &
  \multicolumn{1}{c}{} &
  \multicolumn{1}{c}{} &
  \multicolumn{1}{c}{} &
  \multicolumn{1}{c}{} &
  \multicolumn{1}{c}{\checkmark} &
  \multicolumn{1}{c}{$\partial$} &
  \multicolumn{1}{c}{}  &
  \multicolumn{1}{c}{} &
  \multicolumn{1}{c}{} \\\hline
  
  \multicolumn{1}{c}{  \cite{shakarami2022resource}} &
  \multicolumn{1}{c}{\checkmark} &
  \multicolumn{1}{c}{} &
  \multicolumn{1}{c}{} &
  \multicolumn{1}{c}{} &
  \multicolumn{1}{c}{} &
  \multicolumn{1}{c}{} &
  \multicolumn{1}{c}{} &
  \multicolumn{1}{c}{\checkmark} &
  \multicolumn{1}{c}{\checkmark}  &
  \multicolumn{1}{c}{\checkmark} &
  \multicolumn{1}{c}{} \\\hline

  \multicolumn{1}{c}{\makecell{Our \\work}} &
  \multicolumn{1}{c}{\checkmark} &
  \multicolumn{1}{c}{\checkmark} &
  \multicolumn{1}{c}{\checkmark} &
  \multicolumn{1}{c}{\checkmark} &
  \multicolumn{1}{c}{\checkmark} &
  \multicolumn{1}{c}{\checkmark} &
  \multicolumn{1}{c}{$\partial$} &
  \multicolumn{1}{c}{\checkmark} &
  \multicolumn{1}{c}{$\partial$}  &
  \multicolumn{1}{c}{$\partial$} &
  \multicolumn{1}{c}{$\partial$} \\\hline

\multicolumn{12}{l}{\begin{tabular}[c]{@{}l@{}} 
RP: Resource Provisioning, AP: Application Placement, RA: Resource Allocation, LB: Load Balancing, ML: Machine Learning\\
\checkmark  denotes detail discussion \\  
$\partial$ denotes partial discussion\end{tabular}}
\end{tabular}%
}
\end{table}

Considering the classifications of RM, we review some of the existing surveys in FC that focused on different aspects of RM.
Focusing on resource scheduling the work in \cite{rahimikhanghah2021resource}, provided a taxonomy based on the objective of improving parameters such as performance, energy efficiency, and resource utilization. However, the study only focused on the literature of scheduling, and only a few studies related to ML were studied. Task scheduling is also one of the important aspects of RM. Furthermore, several review papers on task scheduling have been conducted \cite{alizadeh2020task,kaur2021systematic,singh2022towards}. The study in \cite{alizadeh2020task}  provided the suitability of algorithms in different environments along with discussing the different metrics considered by different scheduling algorithms. Similarly, \cite{kaur2021systematic} also provided scheduling strategies such as stochastic, deterministic, and hybrid. In addition, this paper provided an overview of different simulation tools used by the studies and also Quality Of Service (QoS) parameters related to task scheduling. Based on the analysis, time (response, execution, completion time, etc.) is one of the crucial parameters considered in the literature along with cost and energy as the second most important features in task scheduling. In addition, another study in \cite{singh2022towards}, focused on task scheduling with the aim to provide a taxonomy in both cloud and FC considering meta-heuristic techniques. Furthermore, the effectiveness of task scheduling algorithms has been evaluated through a detailed set of criteria such as scheduling objectives, type of mechanism/tasks, metrics, workload generation, testing environment, as well as other important issues like convergence, statistical, and complexity analysis. However, the study did not provide enough details on the classification of scheduling  methods.

Furthermore, considering the studies on application placement, the survey in \cite{nayeri2021application}, focused on the artificial intelligence (AI) perspective of application placement in FC. The work classified the AI approaches used for application placement into evolutionary, machine learning, and combinatorial algorithms.
Another review study in \cite{mahmud2020application}, focused on the different aspects of application management such as: How applications are composed, placed, and maintained. The study also provided taxonomies and covers the different dimensions of application management such as architecture, placement, and maintenance.

The survey in \cite{shakarami2022resource}, focused on the resource provisioning aspect of FC. The aim of their study was to provide a detailed classification of heuristic\textbackslash meta-heuristics, framework-based, ML, game theoretic, and model-based approaches. Furthermore, the future directions were classified into resource migration, context awareness, uncertainties, resource elasticity, and resource performance. However, this survey only covers the resource provisioning dimension of RM 

Most of the literature on RM neglects the importance of ML methods in FC. Therefore, the authors in \cite{samann2021fog}, provided an overview on the importance of ML literature in FC. The authors classify ML approaches into three perspectives of FC, such as; paradigm enhancer, application solution, and security and privacy-preserving. 
However, the survey did not cover the different dimensions of RM. In addition, the paper only considered very few studies of ML in the FC.
In addition, the survey in \cite{abdulkareem2019review} also provided an overview on the role of ML in FC paradigms. The work investigated the use of ML-based techniques in RM, security, and accuracy aspects of FC and edge computing. The authors mainly focused on the computing, decision making, resource provisioning, and delay prediction aspects of RM. 

Although, ML-based approaches for RM have significant importance in FC. However, besides its importance, it is difficult to evaluate these intelligent solutions in the real world due to the heterogeneity of FC and the lack of test beds \cite{samann2021fog}. The existing literature does not focus on the contribution of ML in RM areas in detail. Furthermore, the reviewed works have either considered little or no literature related to ML-based approaches used to achieve RM objectives. For the aforementioned reasons, we focus on the role of ML in various dimensions of RM in FC. 

\section{Resource Management}

RM is one of the important issues in FC. As mentioned in earlier sections, we classified the RM into six sub-dimension such as resource provisioning, application placement, scheduling, resource allocation, task offloading, and load balancing. Each of these dimensions of RM is explained along with an overview of ML works in detail in the following sections.     


\subsection{Resource provisioning}

Resource provisioning can be defined as, the provisioning of resources to the user's requests. Provisioning models can be mainly categorized into user-centric, dynamic, and static \cite{shakarami2022resource}. Based on categorization in \cite{shakarami2022resource}, user-centric models; the resources are provisioned based on the user demands. However, such a model results in the cost of service customizations along with underutilized services. In dynamic models, the resource requirement for the fluctuating workload is fulfilled dynamically through auto-scaling. However, this may lead to over or under-provisioning of resources due to accurate predictions of resource demands. Thus, resulting in cost or volition of service level agreements respectively. Therefore, continuous monitoring; ensures the availability of resources when required. Furthermore, in the static model, the resources are provisioned statically for a certain period of time. However, this model results in wastage of resources if not properly utilized. The dynamic provisioning (i.e auto-scaling), can help in scaling in and scaling out, to provision appropriate resources to the fluctuating demand. The dynamic provisioning can be achieved by adopting different types of policies such as reactive, proactive, and hybrid \cite{ghobaei2020resource}. Reactive is the real-time provisioning of resources to the user demands based on the system's current state. Whereas, the proactive policy is estimating future user demands for the resources. However, in hybrid policy, both reactive and proactive policies are utilized to scale out (release excess resources) and scale in (add more resources). The proactive policy mostly depends on predictive approaches. Therefore, to predict appropriate resources for future user demand, ML-based approaches such as neural networks (NN), reinforcement learning (RL), deep learning (DL), and others can play a vital role in improving resource provisioning accuracy \cite{shakarami2022resource} \cite{guevara2020classification}. The summary of ML works in resource provisioning can be seen in Table \ref{Resource_ Provisioning}

\begin{table}[ht]
\centering
\caption{Resource Provisioning}
\label{Resource_ Provisioning}
\def\arraystretch{2}
\resizebox{\textwidth}{!}{%
\begin{tabular}{cccccccccc}
\hline
\multirow{2}{*}{\textbf{Ref.}} &
  \multicolumn{5}{c}{\textbf{Objective}} &
  \multirow{2}{*}{\textbf{Method}} &
  \multirow{2}{*}{\textbf{Tool}} &
  \multirow{2}{*}{\textbf{Dataset}} &
  \multirow{2}{*}{\textbf{Comparison}} \\ \cline{2-6}
 & \textbf{Cost} & \textbf{Energy} & \textbf{Latency} & \textbf{\makecell{Execution \\ Time}} & 
 \textbf{\makecell{Resource \\ utilization}} &  &  &  \\ \hline
 
\cite{etemadi2021learning} & \checkmark & \checkmark & \checkmark &  & & \makecell{Nonlinear Auto-\\Regressive (NAR),\\ Hidden Markov \\ Model (HMM)} & \makecell{iFogSim}  &  \makecell{Real}& \makecell{ENORM, HSAM}\\ \hline

\cite{etemadi2021cost} &\checkmark &   &\checkmark &  & \checkmark & \makecell{Deep learning based \\ recurrent neural\\ network (RNN)} &iFogSim & \makecell{Synthetic, \\Real} & \makecell{ Multiple linear \\
regression (MLR), \\ Unmanned aerial \\ vehicle (UAV) }\\ \hline
 
\cite{faraji2022self} & \checkmark & \checkmark  & \checkmark  & &  & \makecell{Deep reinforcement \\ learning (DRL)}  & iFogSim & Synthetic & \makecell{ Learning Automata, \\ RL} \\ \hline
 
\cite{pg2022green} & &\checkmark & \checkmark&  &  &\makecell{Autoregressive \\integrated moving \\average (ARIMA)}  & N/A& N/A &N/A\\ \hline

\cite{faraji2021proactive}& \checkmark& &\checkmark & &  &\makecell{Linear regression\\ (LR), ARIMA and, \\learning automata (LA)}  & Java & Synthetic & \makecell{Greedy FogPlan \\method, Optimization \\ based Router method} \\ \hline

\cite{santos2021resource}& \checkmark & \checkmark & &  &  &\makecell{Deep reinforcement \\ learning (DRL)}  & OpenAi gym & Synthetic & \makecell{Mixed-integer \\ linear programming \\ (MILP)}\\ \hline

\cite{murthy2021double}&  &  &    &  \checkmark & \checkmark & \makecell{Double-state-\\temporal difference \\learning}  &iFogSim &Synthetic &Bayesian learning \\ \hline

\cite{abdullah2020predictive}&&&\checkmark&&  &\makecell{Rule based, \\ Decision tree\\ regression (DTR)}  & N/A& \makecell{Synthetic, \\Real} &\makecell{Support Vector \\Regression and fast\\ Fourier transform (FFT)}\\ \hline

\end{tabular}%
}
\end{table}

\textbf{Overview of ML-Based Approaches:}

Several ML-based techniques have been used for tackling the issues of resource provisioning. Most of the studies resolve the issue of resource provisioning while adopting the dynamic proactive policy. For instance, the work in \cite{etemadi2021learning}, addresses resource provisioning through learning-based solutions. The authors use nonlinear auto-regressive (NAR) neural networks for predictions and the Hidden Markov Model (HMM) on the decisions of provisioning resources for serving workloads of IoTs. The matrices considered for the evaluation of the proposed approach are cost, energy, and delay. However, the work also suggests the use of the long short-term memory (LSTM) model as a prediction in the analysis phase and federated learning as a decision-maker in the planning phase will provide better results. The study in \cite{etemadi2021cost}, considers the dynamic resource provisioning model. The work proposed a cost-efficient auto-scaling mechanism using a DL-based recurrent neural network (RNN) for taking decisions such as (scale up/scale down/no action). The proposed model is evaluated considering the matrices such as cost, CPU utilization, delay violation, and network usage. However, it would be interesting to test the proposed method in more real-time case studies and integrate blockchain-based systems and renewable energy into resource auto-scaling.
The work in \cite{faraji2022self}, provides deep reinforcement learning (DRL) solution to minimize response time, cost, and energy consumption. Similarly, the works in \cite{pg2022green,faraji2021proactive}, considers resource provisioning as a time series problem. Linear regression (LR) and ARIMA models are used to predict incoming future workloads and deploy needed resources for fluctuating workloads. Furthermore, RL-based learning automata (LA) method is used for provisioning decisions\cite{faraji2021proactive}. In the proposed approach, the main focus is to reduce energy consumption without violating SLAs.


Furthermore, adopting dynamic reactive policies, the study in \cite{santos2021resource}, uses DRL to achieve cost-effective real-time resource provisioning for energy-efficient FC. However, the work may not be able to perform well with changing resource demands. Furthermore, the proposed method requires a large amount of data to work well and may result in an excessive amount of time. Therefore it is important to consider the issue of lack of information. The study in \cite{murthy2021double}, considers resource provisioning issues under an uncertain FC environment and adopts a reactive policy. The proposed work uses RL based double-state-temporal difference learning method to provide better resource utilization and accuracy along with lower execution time. The study in \cite{abdullah2020predictive}, considers reactive and predictive policy for auto-scaling of microservices in a FC environment. A rule-based reactive policy is adopted at the initial stage for auto-scaling and creating a training dataset. Whereas, decision tree regression (DTR) based predictive policy is used for auto-scaling in the later phase. The combination of reactive and predictive policy approaches minimizes the long response time compared to simple relative auto-scaling methods.


\subsection{Application Placement}
Application placement can be defined as the optimal placement of applications among different layers of FC nodes. Whereas, optimal placement of applications is the mapping of application modules to the available resources in the best possible way to achieve better Quality of Experience (QoE) and QoS matrices. \cite{ghobaei2020resource,nayeri2021application}. Application placement has several components, resource types, placement strategies, orientations, mapping techniques, placement controller, and placement metrics \cite{mahmud2020application,kansal2022classification}.
\begin{table}[ht]
\centering
\caption{Application Placement Approaches}
\label{Application_Placement}
\def\arraystretch{2}
\resizebox{\textwidth}{!}{%
\begin{tabular}{ccccccccc}
\hline
\multirow{2}{*}{\textbf{Ref.}} &
  \multicolumn{4}{c}{\textbf{Objective Matrices}} &
  \multirow{2}{*}{\textbf{Method}} &
  \multirow{2}{*}{\textbf{Tool}} &
  \multirow{2}{*}{\textbf{Dataset}} &
  \multirow{2}{*}{\textbf{Comparison}} \\ \cline{2-5}
 & \textbf{Cost} & \textbf{Energy} & \textbf{Latency} & \textbf{\makecell{Execution \\Time}} &   &  &  \\ \hline

\cite{goudarzi2021distributed}& \checkmark&\checkmark&&\checkmark&\makecell{DRL, RNN}& Python & Real&\makecell{Double-DQN,\\ Proximal Policy\\
Optimization (PPO)}\\\hline

\cite{sami2021demand} & \checkmark&&\checkmark&\checkmark&\makecell{ MDP, Deep Q-\\Network (DQN)} &Python & Real & \makecell{Heuristic based \\ approaches} \\\hline

\cite{eyckerman2022application} &\checkmark&\checkmark&\checkmark& &\makecell{Multi-Objective \\Reinforcement Learning\\ (MORL)} & N/A &Synthetic&\makecell{Non-dominated Sorting\\ Genetic Algorithm II\\ (NSGA-II)} \\\hline

\cite{poltronieri2021reinforcement} & &&&\checkmark&\makecell{Deep Q-Network\\ (DQN)}&Python&Synthetic& \makecell{Quantum Particle \\Swarm
Optimization \\ (QPSO)}\\\hline

 

\end{tabular}%
}
\end{table}
Applications are placed on different nodes having different levels of resources. These resources can be categorized into virtual machines (VMs), containers, and bare metals. The placement algorithms depend on the fluctuating workloads. Hence, two types of placement strategies can be adopted such as static and dynamic \cite{ghobaei2020resource}. In static, applications are placed only once using application placement algorithms and are kept running. Whereas, the execution of placement algorithms depends on the fluctuation of workload. Therefore, in dynamic, instances of applications are either added, terminated or the whole application is migrated due to mobility reasons. One of the important factors is the distributed architecture of nodes in FC. Therefore, it is not possible to avoid the orientations of different nodes in application placement decisions, as it can affect the QoS (e.g. communication delay.). Similarly, several mapping policies can be adopted based on the objective metrics such as priority, optimization, and multi-objective trade-off \cite{mahmud2020application}. Where priority is placing an application on a certain node and optimization is the placement of application based on minimization or maximization of a certain objective metric, lastly, the placement considering multiple objectives metrics adopts a multi-objective trade-off policy. The placement controller works as a manager for controlling the placement activities and can be categorized into centralized and decentralized. In centralized, the responsibility of managing the application is assigned to the commonly accessible node. Whereas in decentralized, the responsibility is managed by the nodes themselves or the brokers. The main objective of application placement is to minimize or maximize the QoE and QoS.

\textbf{Overview of ML-Based Approaches}

The work in \cite{goudarzi2021distributed}, uses the DRL technique for experience-based application placement. Applications can be modeled as Directed Acyclic Graphs (DAGs) with varying numbers of tasks and dependency models. Therefore, to represent the scenarios of IoTs generating heterogeneous DAGS, the authors generated the synthetic DAG datasets with different preferences. The main focus of the work is application placement however, they also consider pre-scheduling for defining task dependencies. The proposed model outperforms other DRL-based techniques in terms of time, energy, and cost by achieving performance gains of 30\%, 11\%, and 24\% respectively. 
Furthermore, the work in \cite{sami2021demand}, proposes an end-to-end architecture and introduced two controllers; an Intelligent Fog Service Scheduler (IFSS) and an Intelligent Fog Service Placement (IFSP) for scheduling and placement of services in distributed fog architecture. The IFSS is used to capture the time and location of environmental change and is based on the R-Learning algorithm proposed in \cite{farhat2020reinforcement} and is responsible for triggering the IFSP agent. An IFSP agent was built using an MDP (Markov Decision Process) and DQN (Deep Q Network) for optimal placement decisions. The proposed IFSP agent performs better in terms of its ability in improving the quality of decisions and reducing decision time. A Multi-Objective Reinforcement Learning (MORL) technique has been proposed in \cite{eyckerman2022application}. The proposed approach places services near to the users to reduce cost along with achieving service reliability by not overloading the fog devices with more services.  


\subsection{Scheduling}

The scheduling problem can be classified into resource scheduling and task scheduling.
In FC literature the terms resource and task scheduling are often used interchangeably \cite{jamil2022resource,kansal2022classification,ghobaei2020resource}. However, resource scheduling is to find the best possible resources for client requests while achieving scheduling goals such as optimal resource utilization. Whereas, task scheduling is the assignment of a set of tasks to available resources in view of the QoS requirements of the tasks. In general, resource scheduling accomplishes the goals of the service providers, and task scheduling provides a better experience to the end-users. 
The scheduling algorithms can be classified into static, dynamic, and hybrid. The static algorithms require prior information (No of task and available resources) for scheduling which is not always a feasible approach. Therefore, dynamic or hybrid approaches are more appropriate and can react more accurately and in real-time to schedule incoming tasks. 
\begin{table}[ht]
\centering
\caption{Resource Scheduling approaches}
\label{Resource_scheduling}
\def\arraystretch{2}
\resizebox{\textwidth}{!}{%
\begin{tabular}{ccccccccc}
\hline
\multirow{2}{*}{\textbf{Ref.}} &
  \multicolumn{4}{c}{\textbf{Objective Matrices}} &
  \multirow{2}{*}{\textbf{Method}} &
  \multirow{2}{*}{\textbf{Tool}} &
  \multirow{2}{*}{\textbf{Dataset}} &
  \multirow{2}{*}{\textbf{Comparison}} \\ \cline{2-5}
 & \textbf{Cost} & \textbf{Energy} & \textbf{Latency} & \textbf{\makecell{Execution \\Time}} &  &  &  \\ \hline

\cite{bhatia2019quantum} &  & \checkmark &  & \checkmark & \makecell{Quantum \\Computing} &  iFogSim & Synthetic & \makecell{RoundRobin, SVM,\\ neural network, MaxMin }\\ \hline

\cite{swarup2021energy}&\checkmark&\checkmark&\checkmark&\checkmark& \makecell{Clipped Double \\Deep Q-learning \\(CDDQL)}&Python& Synthetic& \makecell{FCFS, random \\scheduling,Q learning \\scheduling (QLS)} \\\hline

\cite{sellami2022energy}&&\checkmark&\checkmark&\checkmark&DRL&Mininet&Synthetic&\makecell{ Deterministic,\\ Random agents}\\\hline

\cite{razaq2022fragmented} & & & \checkmark & & Q-learning & Python & Synthetic & \makecell{N/A}  \\\hline

\cite{nair2022reinforcement}& & &\checkmark& &  \makecell{Brain-inspired \\ rescheduling decision-\\making (BIRD) } & N/A &Real& \makecell{shortest\\ job first (SJF), FCFS, greedy, earliest\\ deadline first (EDF)}\\ \hline

\end{tabular}%
}
\end{table}

\textbf{Overview of Machine Learning-based Approaches:}

Quantum Computing inspired-Neural Network (QCi-NN) method is proposed for real-time scheduling in  \cite{bhatia2019quantum}. The proposed method is composed of three layers, each layer having its own functionality such as monitoring, learning, and predicting. In the learning phase, the QCi-NN observes the temporal behavior (arrival time, capacity required, and execution time) of the tasks assigned to specific nodes and stores them in a temporal database. The information stored in the database is used for predicting the optimal fog node for the tasks in the last phase. The proposed technique provides better results in terms of task completion time and energy consumed. However, the work does not consider task dependencies. Although, the proposed approach is for real-time scheduling, but the constraints used are not real-time \cite{Kim2021Dec}.      
Furthermore, it is important to consider rescheduling tasks to avoid long waiting times. Therefore, a DRL-based Clipped Double Deep Q-learning (CDDQL) is proposed in \cite{swarup2021energy}. The proposed task scheduling problem minimizes latency, energy, and cost. This algorithm is used by a parallel scheduler which is equipped in each fog node. Therefore, a parallel queuing approach was used to decrease the waiting time for tasks and optimize resource allocation. Also, regarding the scheduling parameters, they allocated the tasks to the servers based on their length and latency. As a result, if a task has a long waiting time, it is rescheduled using the dual queue approach. Similarly the work in \cite{nair2022reinforcement}, also focuses on the rescheduling of the preemptive task in FC to guarantee the QoS requirement of the tasks. An actor-critic RL-based Brain-Inspired Rescheduling Decision-making (BIRD) algorithm is proposed to achieve the deadline requirement of the user task through rescheduling.
An energy-aware task scheduling approach in SDN enabled FC environment is proposed in \cite{sellami2022energy}. A DRL-based approach under energy constrained is formulated for dynamic task assignment and scheduling. The proposed approach ensures less energy consumption along with reduced latency. The work in \cite{razaq2022fragmented}, proposes a RL based Q-learning approach for latency and privacy-sensitive task scheduling based on QoS requirements such as latency and security. The tasks are grouped into low, medium, and high based on security requirement level. The proposed approach performs better with fewer violations in a delay while ensuring security in allocating tasks.   

\subsection{Resource Allocation}
In FC, resource allocation is allocating a set of tasks having different QoS requirements to a set of heterogeneous fog nodes in a way that it helps in achieving better response time, resource utilization, less energy consumption, cost  \cite{jamil2022resource,kansal2022classification,alsadie2022resource}. User devices, fog nodes, and cloud servers are multiple distributed heterogeneous entities in fog networks. Due dynamic and dense heterogeneous nature of the fog network, resource allocation is challenging and is considered an NP-hard problem \cite{kansal2022classification}. However, in order to efficiently allocate resources to the incoming users' tasks, it is required to consider some unavoidable factors in the FC environment, such as the heterogeneity of applications, stochastic workload environment, and mobility \cite{jamil2022resource}. Applications can be real-time or delay tolerant. The arrival rate and the duration of the tasks are stochastic and fluctuate so the requirements for the resources also change. Mobility also in FC computing can degrade user experience if not properly managed. Mobility introduces several challenges by introducing numerous constraints such as speed, time, and distance.
Several methods have been proposed for resource allocation while considering the above-mentioned factors. These methods can be broadly categorized as auction-based and optimization-based. This broad categorization contains sub-categories such as game theory, heuristics, meta-heuristics, fuzzy, and ML-based solutions \cite{ghobaei2020resource}.

\begin{table}[ht]
\centering
\caption{Resource Allocation Approaches}
\label{Resource_Allocation}
\def\arraystretch{2}
\resizebox{\textwidth}{!}{%
\begin{tabular}{cccccccccccc}
\hline
\multirow{2}{*}{\textbf{Ref.}} &
  \multicolumn{7}{c}{\textbf{Objective Matrices}} &
  \multirow{2}{*}{\textbf{Method}} &
  \multirow{2}{*}{\textbf{Tool}} &
  \multirow{2}{*}{\textbf{Dataset}} &
  \multirow{2}{*}{\textbf{Comparison}} \\ \cline{2-8}
 & \textbf{Cost} & \textbf{Energy} & \textbf{Latency} & \textbf{\makecell{Execution \\Time}} &\textbf{\makecell{Resource\\ Utilization}} &\textbf{Makspan}&\textbf{Throughput} &  & & \\ \hline
 
\cite{talaat2022effective} &  &   & & & \checkmark&\checkmark& & Deep-RL, PNN & iFogSim, Python & Real & \makecell{Round Robin (RR),\\Weighted Round Robin\\ (WRR), Least \\Connection (LC)}\\ \hline

\cite{zhang2022computation} & \checkmark & \checkmark  &\checkmark & & &  & & \makecell{Deep deterministic\\ policy gradient\\ (DDPG)}& N/A & Synthetic & \makecell{Centralized DDPG,\\Local computing}\\ \hline

\cite{talaat2022effective2}&&&&&\checkmark&\checkmark&& \makecell{Optimized RL}&Python&Real&\makecell{Round Robin (RR)\\Weighted Round Robin\\ (WRR) Least \\Connection (LC)}\\\hline

\cite{lakhan2022efficient} &\checkmark& \checkmark&\checkmark&\checkmark&&&&\makecell{Deep-learning-\\Q-network \\Based (DQB)}& Python-Ruby-Perl & Synthetic &\makecell{Heterogeneous earliest\\ finish time (HEFT),\\min–max, Markov\\ decision process}\\\hline

\cite{tan2022resource}&\checkmark&\checkmark&&&&&\checkmark&\makecell{Q-Learning\\ and Deep Q \\Network (DQN)}&Matlab&Synthetic&\makecell{Greedy algorithm,\\ random allocation \\algorithm}\\\hline

\cite{khumalo2021reinforcement} & \checkmark&&\checkmark&&\checkmark&&&Q-learning& 5G K-SimNet& Synthetic&\makecell{SARSA \\ Monte Carlo} \\\hline

\cite{santos2021reinforcement} & \checkmark &&&&&&&  Q-learning & Python &Synthetic&MILP \\ \hline

\cite{naha2021multiple} &&\checkmark&\checkmark&\checkmark&&&& \makecell{Multiple linear\\ regression}&CloudSim& Synthetic&\makecell{DQN algorithm,\\ greedy algorithm,\\random algorithm} \\\hline



\end{tabular}%
}
\end{table}

\textbf{Overview of ML-Based Approaches:}

One of the key objectives of resource allocation is to reduce the completion time. Therefore, the work in \cite{talaat2022effective}, proposed an effective prediction-based resource allocation method (EPRAM), by using a DRL and Probabilistic Neural network (PNN). The proposed method uses a DRL for resource allocation decisions, however, the target destination decisions are based on the PNN prediction algorithm.
The proposed approach is effective in achieving minimum makespan while increasing resource utilization. A hybrid Optimized RL (ORL) method is used in another study in \cite{talaat2022effective2}, to achieve minimum makespan along with increased resource utilization. In the proposed work, Particle Swarm Optimization (PSO) has been used to optimize the parameters of RL and RL for resource allocation. 
Considering the factor of mobility, a DRL-based Deep-learning-Q-network Based (DQB) method is proposed for resource allocation in an SDN-enabled FC paradigm \cite{lakhan2022efficient}. 
The proposed algorithm adopts a mobility-aware policy to reduce costs in terms of energy and execution time. 
Furthermore, for the issue of resource allocation in Fog Radio Access Networks (F-RAN), the work in \cite{tan2022resource}, adopts RL-based Q-Learning and Deep Q Network (DQN) for optimizing resource allocation. The proposed optimization algorithm increases the throughput while achieving constraints such as signal-to-noise ratio (SNR), bandwidth, and energy. Similarly, another work in the 5G F-RAN environment proposes Q-learning for dynamic and autonomous resource allocation \cite{khumalo2021reinforcement}.

The proposed approach increases resource utilization along with reduced cost and latency. Another approach in  \cite{zhang2022computation}, uses federated DRL-based deep deterministic policy gradient (DDPG) policy to offload and allocate computational resources in F-RAN. The proposed policy effectively reduces cost, energy, and delay.
Furthermore, considering the factor of applications the work in \cite{santos2021reinforcement} focuses on service function chain (SFC) allocation in fog computing. Where RL agent is developed to learn the best decision for resource allocation. The proposed approach results in a higher acceptance rate and low cost compared to the MILP model. Energy-aware multiple linear regression-based resource allocation is proposed in another work in \cite{naha2021multiple}. the proposed approach investigates the trade-off between energy consumption and execution time and minimizes the delay and response time significantly.


\subsection{Task offloading}
Offloading in a computing environment can be defined as releasing resource-hungry tasks from resource-constrained devices and assigning them to resource-rich devices \cite{ghobaei2020resource}. The question arises on what, where, and how to offload. The offload decision on what to offload can be data, computation, or application \cite{bukhari2022intelligent}. However, to answer where to offload, the offloading policy can be no offload or offload task either horizontally at the same level or vertically to fog or cloud. For instance, horizontal offloading can be device-to-device or fog-to-fog whereas vertical offloading refers to device-to-fog or fog-to-cloud \cite{ahvar2021next}. Several techniques have been proposed in the literature to adopt these policies such as AI, approximation, mathematical modeling, heuristics, meta-heuristics, game theoretic, and fuzzy-based approaches \cite{guevara2020classification}, in order to address several challenges.

\textbf{Overview of ML-Based Approaches:}

Several studies based on RL have been conducted recently for efficient task offloading. The work in \cite{jiang2021reinforcement}, achieves a higher utility score along with lower latency and energy consumption by adopting a hybrid policy of no-offload, device-to-fog, or fog-to-cloud. A centralized dueling DQN (DDQN) approach is used to obtain the most suitable computational offloading policy and a decentralized DQN to optimize resource allocation at the fog level. Another study in \cite{shi2021deep}, utilizes DRL to achieve high utility and less delay in a heterogeneous vehicular fog computing environment. The proposed work adopted the policy of fog-to-fog partial computational offloading by adding a centralized learning approach in the base station (BS). 

\begin{table}[ht]
\centering
\caption{Task offloading}
\label{Task_offloading}
\def\arraystretch{2}
\resizebox{\textwidth}{!}{%
\begin{tabular}{ccccccccc}
\hline
\multirow{2}{*}{\textbf{Ref.}} &
  \multicolumn{4}{c}{\textbf{Objective Matrices}} &
  \multirow{2}{*}{\textbf{Method}} &
  \multirow{2}{*}{\textbf{Tool}} &
  \multirow{2}{*}{\textbf{Dataset}} &
  \multirow{2}{*}{\textbf{Comparison}} \\ \cline{2-5}
 & \textbf{Cost} & \textbf{Energy} & \textbf{Latency} & \textbf{\makecell{Execution\\Time}}  &  &  &  \\ \hline

 
  \cite{jiang2021reinforcement}&  & \checkmark&\checkmark &  &   \makecell{Dueling deep Q-\\network (DDQN), Deep\\ Q-Network (DQN)}  & \makecell{Python, \\Adam optimizer}& Synthetic & \makecell{Random task offloading \\ (RO), non-D2D offloading\\ (ND2D), non-pre-processing\\ offloading (NP)} \\ \hline
 
 \cite{shi2021deep}& &&\checkmark &  &\makecell{DRL}  & N/A & Synthetic & \makecell{Random-Based Partial \\ Offloading (RBPO), Greedy-\\Based Partial Offloading\\ (GBPO)} \\ \hline
 
 \cite{ren2020deep}& &\checkmark& &  & \makecell{Multi-agent DRL} &N/A &Synthetic & \makecell{Genetic algorithm, Exhaustive\\ search, Priority based,\\ Random based, Greedy\\ algorithm based,\\ Single-agent DRL} \\ \hline

 \cite{jazayeri2021autonomous}& \checkmark&\checkmark&\checkmark &  &DRL  &iFogSim &Synthetic & \makecell{No-Offload, first\\ fit (FF), and ASDEO  } \\ \hline
 
 \cite{sarkar2022deep}& \checkmark& \checkmark&\checkmark &&    \makecell{Deep neural \\network (DNN)}& Python& Synthetic &\makecell{Random Offloading Approach\\ (ROA),Greedy-based \\Offloading Approach (GOA)} \\ \hline
 
 \cite{yang2020online}& && \checkmark &&    \makecell{Restless multi-armed \\ bandit (RMAB)}& N/A& Synthetic& \makecell{Utility-based learning (UL),\\ adaptive learning task\\ offloading (ALTO), regenerative\\ cycle algorithm (RCA),  \\Deterministic sequencing of\\ exploration and exploitation
(DSEE) } \\\hline
 
 \cite{cho2021learning}& \checkmark &\checkmark&\checkmark &  &\makecell{Adversarial multi-armed \\ bandit theory}  & Matlab &\makecell{Synthetic,\\ Real} & \makecell{Implicit algorithms \\ with bandit half and\\ full-feedback} \\ \hline
 
 \cite{gao2021integration}& &\checkmark&\checkmark&  & Bandit learning & N/A& Synthetic & \makecell{LAGO-UCBT,\\ LAGO-$\varepsilon$-greedy,\\ and LAGO-NConfR}\\ \hline
 
\cite{suryadevara2021energy}& &\checkmark&\checkmark &    & Decision Tree (DT)&iFogSim, Weka& Synthetic & \makecell{SVM, k-NN,\\ and Naïve Bayes} \\ \hline

\cite{bukhari2022intelligent}&\checkmark &\checkmark&\checkmark&  & Logistic regression (LR) & Python, Matlab &Real& \makecell{k-NN,\\ Naive Bayes, Decision Tree,\\ SVM and MLP}\\ \hline

\end{tabular}%
}
\end{table}
Furthermore, the study in \cite{ren2020deep}, focuses on computational offloading in fog-enabled industrial IoTs. The work uses multi-agent DRL for the decision on the selection of an appropriate fog access point (F-AP). The proposed work adopted the policy of device-to-fog or fog-to-cloud to reduce the energy consumption of users and fog devices. A Monitor-analysis-plan-execution (MAPE) controller has been used for determining the computational offloading policy in \cite{jazayeri2021autonomous}. Greedy auto-scale DRL (GASDEO) based approach is used for finding the best available destination for computational offloading. In addition, the work also provides an auto-scalable  strategy when the demands increase. The proposed work achieves reduced energy, cost, and delay along with better utilization of resources. Another study on computational offloading in a heterogeneous fog environment has been proposed in \cite{sarkar2022deep}, for reducing energy consumption, cost, and delay. The  proposed work uses DRL based binary offloading policy where it considers multiple parallel deep neural networks (DNNs) for offloading decisions. These decisions are further stored for future testing and training of DNNs. The proposed policy greatly reduces the computational time of offloading decisions.

Multi-armed Bandit (MAB) based decentralized learning approach is considered in \cite{yang2020online,cho2021learning,gao2021integration}. The study in  \cite{yang2020online}, applies restless MAB (RMAB) while adopting an incremental sequence policy for exploitation and exploration for multi-user offloading in a fog environment for reducing the average latency. 
Similarly, \cite{cho2021learning}, proposes a learning algorithm based on adversarial MAB theory for selecting less complex vehicular fog nodes for task offloading. The proposed algorithm achieves a better exploration and exploitation balance along with minimizing cost in terms of latency and energy. 
Similarly, the algorithm in \cite{gao2021integration}, obtained an optimal trade-off between latency and energy under energy constraints.

The work in \cite{suryadevara2021energy}, reduces fog gateway latency and energy through an ML-based data classification method. Based on data classifications results, offloading policy such as device-to-fog or to-cloud is adopted. The work in \cite{bukhari2022intelligent}, proposes a task offloading model based on logistic regression. The proposed work is efficient in predicting to adopt the offloading policies such as no-offload, device-to-fog, and fog-to-cloud offload to achieve less latency, energy, and cost.


\subsection{Load Balancing}
The appropriate Load Balancing (LB) mechanisms help in reducing response time and energy consumption while increasing throughput \cite{kashani2022load}. The LB architecture can be broadly categorized into centralized, decentralized, and semi-decentralized \cite{ghobaei2020resource}. 
\begin{table}[ht]
\centering
\caption{Load Balancing Approaches}
\label{Load_Balancing}
\def\arraystretch{2}
\resizebox{\textwidth}{!}{%
\begin{tabular}{cccccccccc}
\hline
\multirow{2}{*}{\textbf{Ref.}} &
  \multicolumn{5}{c}{\textbf{Objective Matrices}} &
  \multirow{2}{*}{\textbf{Method}} &
  \multirow{2}{*}{\textbf{Tool}} &
  \multirow{2}{*}{\textbf{Dataset}} &
  \multirow{2}{*}{\textbf{Comparison}} \\ \cline{2-6}
 & \textbf{Cost} & \textbf{Energy} & \textbf{Throughput} & \textbf{\makecell{\makecell{Response\\Time}}} &\textbf{\makecell{Resource\\Utilization}}&  &  &  \\ \hline

\cite{albalawi2022load}& &\checkmark& \checkmark   &\checkmark&\checkmark& \makecell{PSO, SVR} & iFogSim  & Synthetic& FCFS, RR \\ \hline

 \cite{hameed2021energy}& &\checkmark& &&\checkmark  & \makecell{Dynamic clustering, \\linear optimization}  &  NS2& Scenarios & Comparisons of scenarios\\ \hline
 
 \cite{liao2020cognitive}& \checkmark &&   &&\checkmark& \makecell{Distributed Q-learning\\ algorithm} & N/A  & Synthetic & \makecell{Centralized Q-learning \\algorithm, maximum and\\ minimum fairness, Markov\\ decision process (MDP)} \\ \hline
 

 \cite{mobasheri2021toward}& &&   &\checkmark&&\makecell{Hierarchical reinforcement\\learning (HRL)} & N/A &Synthetic & \makecell{RL with fixed \\threshold } \\ \hline
 
 \cite{talaat2020load}&\checkmark &\checkmark&&\checkmark&\checkmark&  Q-learning& Matlab& Real &\makecell{Least connection (LC)\\Round robin (RR)\\Weighted round robin (WRR)} \\\hline
\end{tabular}%
}
\end{table}
In a centralized approach, a central node acts as a global LB node, where it contains information on all the available resources in the network. Whereas, the decentralized approach has more than one LB node. Usually, this type of architecture is adopted when you have multiple clusters in a network. Finally, in a semi-decentralized approach, multiple nodes have local Load balancers, where a global load balancer is used to communicate information to and from local load balancers. However, due to the heterogeneous nature of FC, the efficient LB mechanism does not only depends on the architectural perspective but also depends on the LB strategies (i.e., static, dynamic, hybrid strategies), algorithms, and parameters considered to achieve certain performance objectives (i.e., minimizing response time, resource utilization, power consumption, cost or improving throughput) \cite{kansal2022classification,kashani2022load,alsadie2022resource,mijuskovic2021resource}).

In static strategy, the workload is distributed among available fog nodes equally irrespective of their storage and processing capabilities. While in dynamic LB the resource information is shared continuously and the load balancer knowledge base is updated regularly with the available resources. Although this strategy helps in achieving less response time and high throughput, it consumes more bandwidth than the static strategy due to its excessive amount of information sharing. Whereas, Hybrid LB combines the behavior of both static and dynamic strategies to provide low bandwidth consumption.


\textbf{Overview of ML-Based Approaches:}
The work in \cite{albalawi2022load}, proposes a decentralized LB architecture and utilizes a hybrid approach (PSOSVR) by combining Particle Swarm Optimization (PSO) with support vector regression (SVR) to randomly initialize the particles in PSO based on the predictions results of SVR to achieve a better solution in less time. The proposed approach performs better than First Come First Serve (FCFS) and Round Robin in terms of response time, energy consumption, throughput, and resource utilization. Furthermore, the work in \cite{hameed2021energy} also uses a hybrid (i.e. Combination of ML and heuristics/meta-heuristics) method by combining dynamic clustering with linear optimization for capacity-based load distribution in a vehicular FC environment. Dynamic clustering is used to select the appropriate cluster head. Whereas, the capacity-based load distribution, distributes the load within a cluster or among the clusters to achieve maximum resource utilization, reduced energy consumption, and less delay. Similarly, in \cite{talaat2020load}, a real-time dynamic resource allocation and LB Optimization Strategy (LBOS) using RL and the genetic algorithm was proposed. The proposed approach ensures the best load balancing level compared to other state-of-the-art approaches; low cost and response time with fewer migrations with maximum resource utilization and low energy consumption.

Another study in \cite{liao2020cognitive}, implements cognitive balancing in FC by introducing an RL-based distributed Q-learning method along with multi-agent consistency theory to reduce cost and increase resource utilization. Similarly, a distributed hierarchical two-level RL (HRL) was adopted in \cite{mobasheri2021toward}, for the fog cooperation problem. 
The proposed work succeeded in achieving a low convergence time at the second level of learning. 


\section{Analysis \& Discussion}

In this section, we analyzed existing literature provided in Table \ref{Resource_ Provisioning} to \ref{Load_Balancing} and here we provide answers to our research questions in Table \ref{research_questions} 

\textbf{How resource management can be classified into sub-areas?}

There are several classifications of RM in the literature \cite{ghobaei2020resource,kansal2022classification,ahvar2021next}. We categorized the dimensions of RM into six areas as shown in Fig \ref{Aspects}. Although this work categorizes RM into sub-areas, efficient RM can only be achieved if all of these dimensions are integrated as one unit. For example, Load balancing cannot be achieved without proper offloading and optimal allocation mechanisms. Similarly, efficient resource provisioning includes resource discovery, scheduling, placement allocation, and load balancing modules. Therefore, RM should be considered holistic, where all these sub-modules contribute in a collaborative manner to achieve efficient RM in FC.    

\textbf{What are the different ML methods used?}

\begin{figure}[ht]
\includegraphics[width=0.5\textwidth]{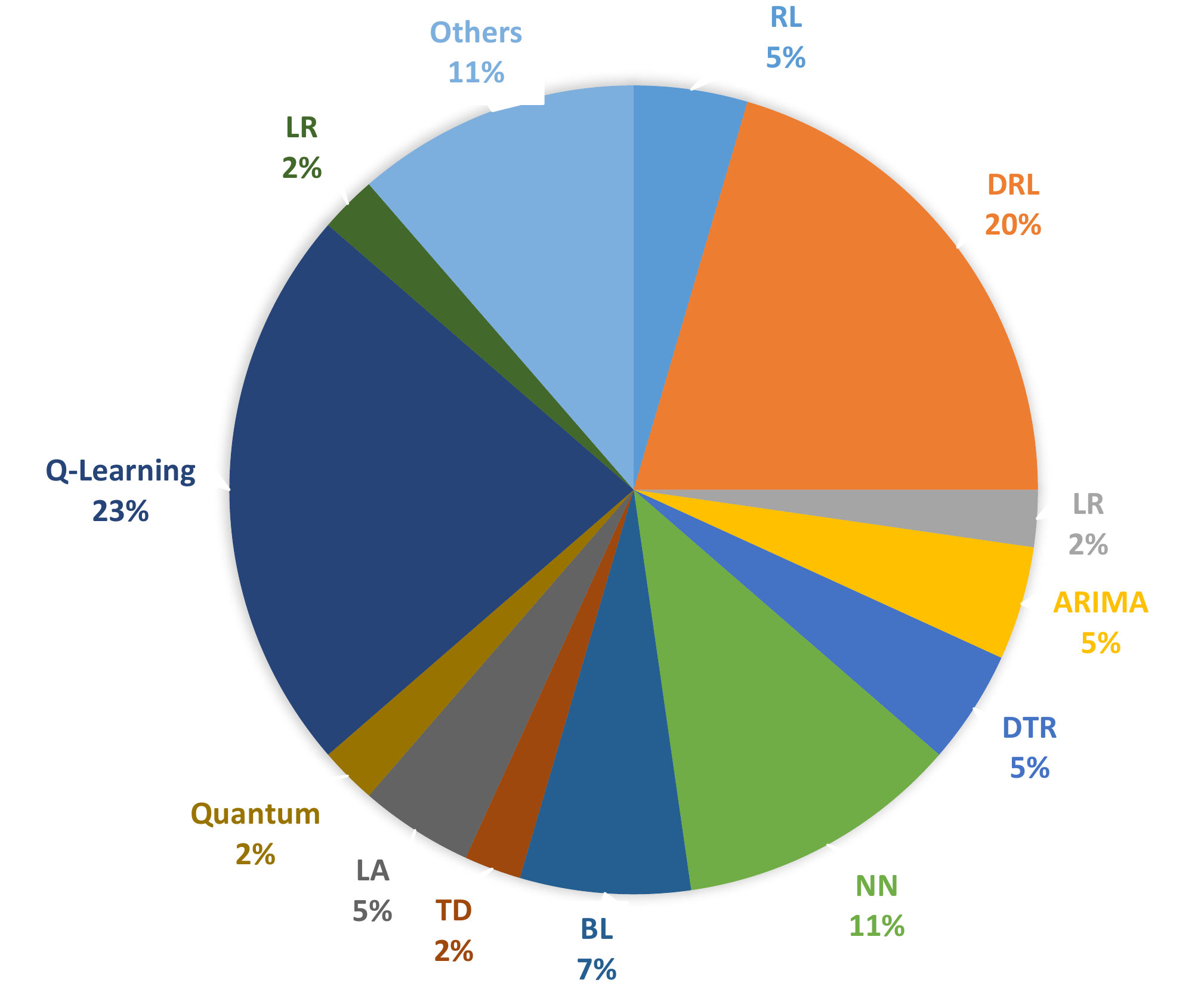}
\caption{Methods used for Evaluation} \label{ML_methods}
\end{figure}

Based on an analysis of different ML approaches (supervised, unsupervised, and RL) used for resolving RM in an FC environment. We found that RL, DL, and NN-based approaches received considerable attention in all areas of RM as shown in Figure \ref{ML_methods}.
Different variations of these approaches are considered such as Q-learning, DRL, RNN, temporal difference learning, DQN, and Bandit Learning. RL is a trial and error approach, where it learns through interacting with the environment to explore and exploit the best possible actions. RL-based approaches are suitable for complex problems and are capable of performing without any prior knowledge. Most of the works that uses RL-based approaches consider the assumption of an uncertain environment, where there is no prior information and RL-based approaches are used to develop information through trials and then take decisions through testing. However, these RL-based approaches provide better results only when knowledge uncertainty decreases and are highly computational due to their huge state and space. To overcome the long computational complexity issue of RL, hybrid approaches such as the combination of RL and DL can be considered effective in an uncertain environment.

Furthermore, we also observed other approaches such as time series analysis (ARIMA) methods for dynamic resource provisioning decisions. In addition, other approaches such as Naive Bayes, Decision Tree (DT), Logistic regression (LR), Decision Tree, Support Vector Machine (SVM), and k-Nearest Neighbours (k-NN) algorithm were applied for the classification of tasks and decision on task offloading. However, the FC paradigm has progressed significantly and the amount of data has increased considerably. Therefore, finding solutions with simple ML-based methods is suggested rather than more complex and computationally intensive approaches. Despite the importance of the time complexity of the algorithms (i.e. The amount of time taken by an algorithm to execute a set of input functions), only a few studies provide the time complexity of their proposed algorithms. The suitability and effectiveness of the approach can be determined if it is compared with the state-of-the-art approaches available in the same category. Most of the work compares their model with outdated approaches or with methods from a different category, which can be questionable and create doubts about the efficiency of the proposed techniques.    

\textbf{What are the different performance matrices considered?}

We analyzed the existing literature to find the most important objective metric considered in various dimensions of RM. For instance, the most important issues considered in scheduling approaches are to reduce cost, mostly in terms of energy along with satisfying latency and time requirements of end user's tasks. Similarly, placing the application near the edge of end users reduces response time resulting in lower cost and energy consumption, and are among the most important objectives considered. We observed that resource utilization is one of the important issues of load balancing. Overloading fog nodes consumes more energy and a long response time results in higher costs and dissatisfaction of end users due to long delays. Most of the reviewed load balancing are multi-objective and consider more than one metric as their objectives such as resource utilization, response time, and cost. Furthermore, in resource allocation, the most common objective considered were cost and energy optimization. Mostly dynamic resource provisioning strategies were adopted to address the issues of under and over-provisioning. Most of the studies considered proactive policies where the resource demands are predicted before actual demand. Cost, energy, and latency were the most common objective metrics. Offloading tasks near the proximity of users result in lower latency compared to the cloud. Furthermore, offloading tasks from resource-limited end-user devices to resource-rich devices available at fog saves a considerable amount of energy at the user end. Therefore, the most common objective metrics considered in the task offloading domain were energy and latency. 

In general the metrics considered in literature can be categorized into two perspectives such as provider's and user's perspectives. The provider's perspective is to have low cost without violating SLAs for achieving better user's experience. The user's perspective is to get better experience in low cost. Therefore, one of the most important challenge is to reduce cost without violating SLAs. In addition to cost, carbon emission is also a challenging issue in the FC environment due to its heterogeneous nature \cite{ahvar2021deca}. Carbon emission as a performance objective is highly neglected in ML works and it has significant importance from the provider's perspective.
Furthermore, the most common non-functional objectives considered in the analyzed RM papers are security and reliability.

\textbf{Which type of data have been considered for evaluation?}

Type of data used in the evaluation stage to test ML-based approaches in RM, it has been observed that mostly the data used for evaluation are based on assumptions and are created synthetically. Furthermore, it is to be noted that almost in every paper, that uses RL based approach considers synthetic data due to the assumption of a lack of information on resource demands. Furthermore, some of the existing approaches were tested using their own test beds data while others with real-world publicly available data traces. Real-world traces used in literature are GWA-T-12 Bitbrains \cite{thegridworkloadsarchive},  Google Cluster Trace 2011-2 \cite{reiss2011google}, MHEALTH dataset \cite{MachineLearningRepository}, Microsoft T-Drive Trajectory Dataset \cite{zheng2011t}, and Chicago Taxi Trips Dataset \cite{Chicagodataportal}. 

\textbf{Which simulation tools have been used?}

\begin{figure}[ht]
\includegraphics[width=0.5\textwidth]{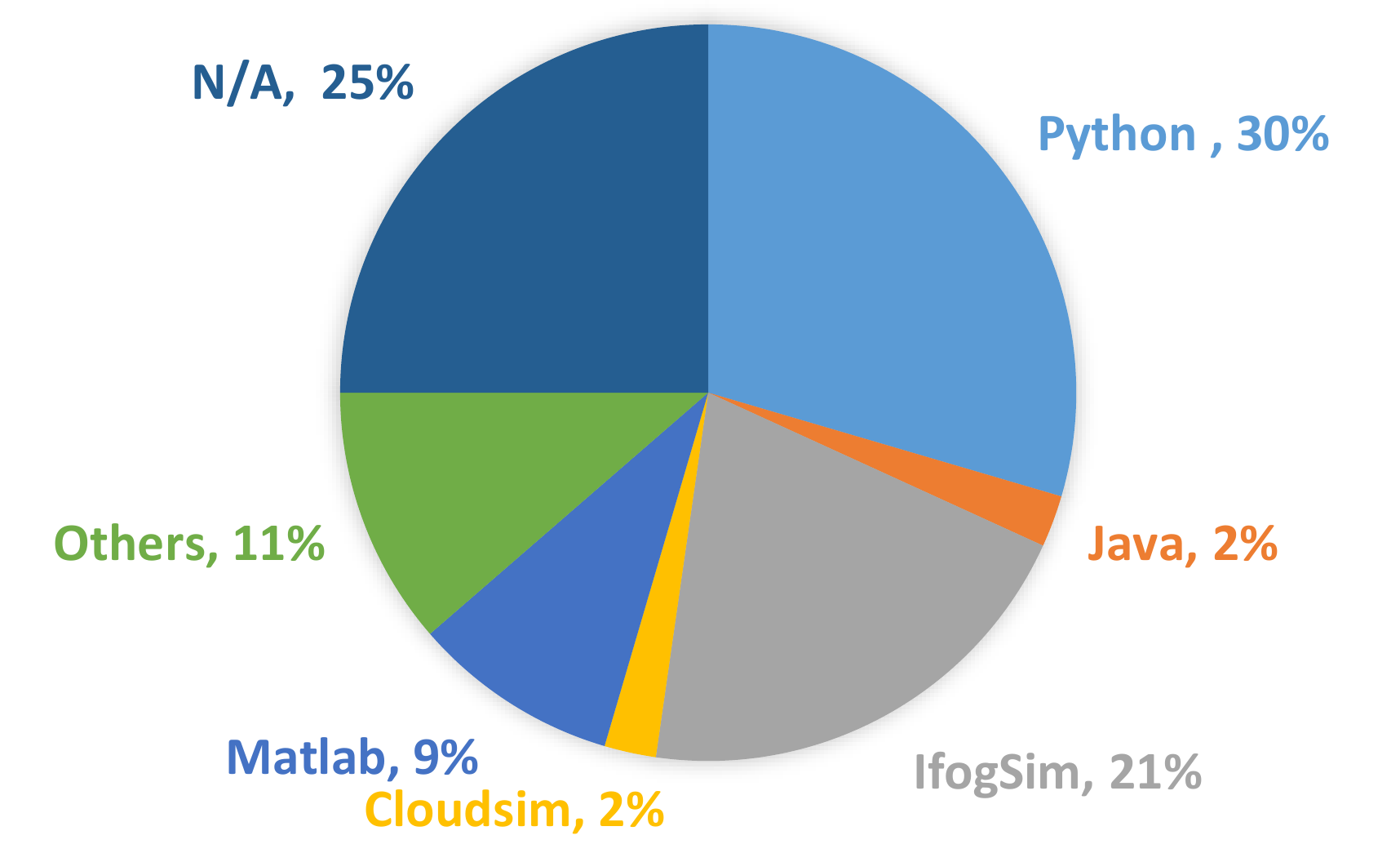}
\caption{Tools used for Evaluation} \label{ML_tools}
\end{figure}

The most common language used for simulation is python and the most common simulator considered for evaluation of experiments is iFogSim as shown in Figure \ref{ML_tools}. It has been observed that mostly when the real datasets were considered for testing, the preference is given to python and java. Whereas, the iFogSim simulation tool was considered when the type of data for evaluation is created synthetically under various assumptions.      

\section{Conclusions}
In this work, a review of ML-based RM approaches has been provided in the FC environment. The RM is classified into six sub-areas resource provisioning, application placement, scheduling, resource allocation, task offloading, and load balancing. The sub-areas are analyzed based on the utilized techniques, tools, data, and metrics. Based on observations, RL approaches are extensively used on the assumption of a lack of information. Most approaches manage the resource through the centralized learning process, whereas few decentralized learning techniques such as bandit learning are utilized to solve RM issues. Furthermore, mobility-aware studies were least focused on in the literature. 

In future work, the focus can be on non-functional challenges (e.g. reliability, scalability, security, etc.) and their architectures. Furthermore, a more detailed classification of RL-based approaches is required due to its extensive use for RM in FC.

\bibliographystyle{unsrt} 
\bibliography{Main}
\end{document}